\documentclass[12pt]{article}

\usepackage{wrapfig}
\usepackage{authblk}
\usepackage{graphicx}
\usepackage{caption}
\usepackage{subcaption}
\usepackage{siunitx}
\usepackage{enumitem}
\usepackage[utf8]{inputenc}
\usepackage[english]{babel}
\usepackage{geometry}
\usepackage{setspace}
\begin{document}


\title{Automatic diffusion path exploration for multivalent battery cathodes using geometrical descriptors}

\date{\today}

\author[1]{\small Felix T. B\"olle}
\author[1]{\small Arghya Bhowmik}
\author[1]{\small Tejs Vegge}
\author[1,*]{\small Juan Maria Garc\'{i}a Lastra}
\author[1,*]{\small Ivano E. Castelli}

\affil[1]{%
Department of Energy Conversion and Storage, Technical University of Denmark, Anker Engelundsvej 411, DK-2800 Kgs. Lyngby, Denmark.}
\affil[ *]{\textit{Email: {jmgla@dtu.dk (JMGL), ivca@dtu.dk (IEC)}}}

\maketitle

\section*{Abstract}
Stable and fast ionic conductors for magnesium cathode materials have the prospect of enabling high energy density batteries beyond current Lithium-ion technologies. So far, only a few candidate materials have been identified leading to data only being scarcely available to the community. 
Here, we present a systematic  study, in the framework of Density Functional Theory, including the estimation of the diffusion barrier for 16 materials through employing Nudged Elastic Band (NEB) calculations. By introducing a path finder algorithm based on the idea of Voronoi tessellations, we show that an estimate of the transition state configuration can be extracted automatically prior to running NEB-calculations. Using geometrical descriptors in combination with a principal component analysis it is possible to further sub-group the diffusion paths. This approach also extends to materials which are not part of the study, making it a viable approach to more efficiently explore crystal structures with distinguishable diffusion characteristics.

\section{Introduction}
Multivalent magnesium-ion batteries hold great promise through increased energy densities. \cite{liang2020current} Finding suitable cathode materials has been difficult since diffusivity is hampered by the increased charge state of magnesium. In terms of stability, the state-of-the art cathode electrode is the Mo$_6$S$_8$ Chevrel phase being stable over hundreds of cycles. \cite{aurbach2000prototype} After its discovery, high cycling performance has been reported for other cathode materials like the thiospinel Ti$_2$S$_4$ structure. \cite{sun2016high} Nevertheless, major drawbacks like the low voltages and poor stability remain. \cite{canepa2017odyssey}\\
Although rather scarce, screening studies have successfully been conducted using Density-Functional Theory (DFT)-based ab-initio molecular dynamics (AIMD) simulations \cite{kahle2020high} as well as DFT in combination with NEB methods \cite{jalem2018bayesian,liu2015spinel,fujimura2013accelerated} to predict the diffusivity in ion insertion materials. We have recently developed a general workflow based on DFT + NEB methods allowing to calculate thermodynamic and kinetic (ionic diffusion) properties automatically across different crystal structures.\cite{bolle2020autonomous} However, one of the main bottlenecks for screening studies based on kinetics descriptors is undoubtedly the computational time it takes to conduct these calculations. In order to explore the possible phase space of ionic conductors more efficiently, one possibility is to find descriptors for NEB paths being able to quantify and describe the distinct diffusion topology of a unique path for a given crystal structure, thus accelerating the discovery process.\\
Previous studies have used few representative structures with distinct diffusion topologies to derive rules for the diffusivity in multivalent magnesium battery materials.  \cite{liu2015spinel, rong2015materials} Geometrical investigations in combination with cheaper computational methods based on Bond Valence Site Energies (BVSE) \cite{katcho2019investigation} or DFT calculations on a few selected structures \cite{gulino2020combined} have been conducted. They discussed the importance of being able to describe the geometry of the transition state along the diffusion path in order to predict kinetic barriers. Compared to the cheaper BVSE method, DFT calculations are able to quantify the effect of ionic and electronic relaxations on the migrating ion needed to get a better estimate of the barrier. \cite{levi2009review,kweon2017structural} \\
Substructure-based analyses of material properties helped in identifying distinct atomic environments found in crystal structures.\cite{pauling1929principles,daams2000atomic}
Voronoi tessellation based descriptors have been successfully employed to predict possible Li sites in electrode insertion materials.\cite{yang2014proposed}
Additionally, a geometric path analysis can be carried out through employing a Voronoi tessellation. \cite{brostow1998voronoi,blatov2004voronoi,pinheiro2013high} While Voronoi tessellations find high symmetry points in space, here we show that these points do not necessarily have to coincide with the transition state geometry found through NEB calculations. Therefore, to solve this issue, we introduce a path finder algorithm extending the ideas of Voronoi tessellations. The algorithm can extract  descriptors forecasting the transition state geometry without prior relaxation of the diffusion path. It solely requires a relaxed supercell leading to a speed-up of approximately three orders of magnitude compared to running the full NEB-calculations. In order to test the algorithm, we conduct a systematic study and determine kinetic barriers for 16 different Mg-ion cathode electrodes. The materials range across different chemical compositions and crystal structures. Using descriptors extracted from the initial and transition state configuration of the path, we discuss how to efficiently guide screening studies when searching for good ionic conducting cathode materials. Additionally, we show that for predicting kinetic barriers the approach of grouping similar diffusion paths is beneficial as it allows to find structure group specific descriptors.

\section{Results and Discussion}
\subsection{Workflow for calculating NEB-barriers}
The workflow for obtaining NEB-barriers has been described in detail in our previous work. \cite{bolle2020autonomous} Here, we only calculate NEB paths that posses a reflection symmetry, \textit{i.e.} all images along the path can be mapped onto a symmetry equivalent image through a reflection operation. Initial Reflective Middle Image-NEB (RMI-NEB) \cite{mathiesen2019r} barriers are calculated for all paths, followed by a climbing image NEB (CIR-NEB) calculation, if the difference in energy between the middle and the initial image is less than 2 eV (for details see Ref. \cite{bolle2020autonomous}).\\

\subsection{Initial pool of materials}
Initially we consider all  Inorganic Crystal Structure Database (ICSD) \cite{ICSD} structures as found in the Materials Project database \cite{jain2013commentary} containing at least three different elements including (1) a metal (M = \{ Ti, V, Cr, Mn, Fe, Co, Ni, Cu, Mo, Sn, Sb, Pb, Nb, Zr \}), (2) an anion (A = \{O, N, S, Se, Te, Cl\}) as well as (3) either Mg or Zn. Zn containing structures are included due to the similar valence and ionic radius of Zn$^{2+}$ and Mg$^{2+}$. These criteria lead to 304 initial structures out of which 105 contain Mg and 199 contain Zn. For the Zn containing structures we replace Zn with Mg. Ensuring that at least one percolating path for the Mg ion exists (\textit{i.e.} ion is not trapped inside the structure and macroscopic ion migration is possible) as well as only considering structures in which the metal atom does not exceed its maximum oxidation state upon fully charging the electrode, reduces the amount of possible structures to 105. We denote a structure a duplicate, if upon exchanging Zn with Mg in the Zn-containing structures, the chemical composition as well as space group match. A NEB path connects two structural configurations with the moving ion residing in the initial and final site, respectively. Geometric interpolation between the initial and final configuration leads to intermediate images along the diffusion path. A path is marked as inaccessible if upon linear interpolation in between the initial and final configuration any of the images contain atom-atom distances below 1 \si{\angstrom}. This leaves us with a total of 77 structures serving as an input for the workflow described in Ref. \cite{bolle2020autonomous}.\\
For estimating the stability against phase separation into more stable counterparts, the energy above the hull is calculated (for details see for instance Ref. \cite{larsen2017atomic}). Upon relaxation of the unit cell with the chosen input parameters, 18 more structures are discarded due to the convex hull threshold of 0.5 eV/atom being exceeded by the fully charged unit cell. The rather high threshold is chosen to account for structures in which only partial charging/discharging is possible due to stability reasons. In these cases, investigating diffusivity at the high and dilute vacancy limit can still give valuable insights on the overall conductivity. \cite{bolle2020autonomous}  Additionally, some of the structures are marked as non-percolating since the relaxation of atomic positions leads to structural changes. 


\begin{figure}[h]
    \centering
    \includegraphics[width=1.\linewidth]{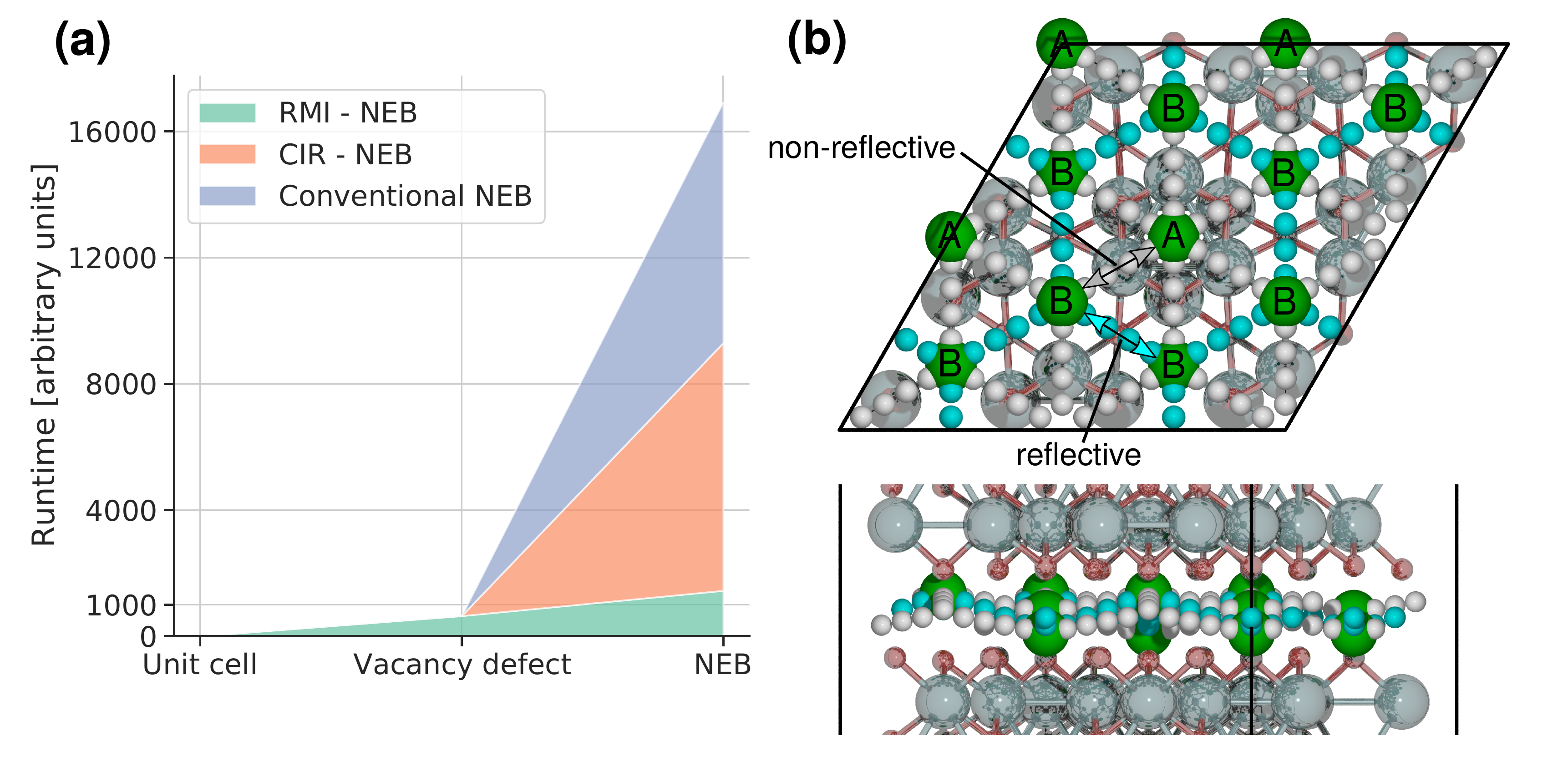}
    \caption{(a) Representative run-time comparison for a single NEB path in a MgTi$_2$O$_4$ spinel structure. All calculations were carried out on the same compute node architecture for better comparability. Relaxing the NEB path is 3-4 orders of magnitudes slower than relaxing the unit cell. (b) The layered Mg$_3$Nb$_6$O$_{11}$ supercell structure as an example of a material consisting of a non-reflective (A-B, light grey) and a reflective (B-B, turquoise) percolating path. The two symmetry inequivalent Mg-ions (green) are labeled as A/B.}
    \label{fig:runtime_mgnbo}
\end{figure}

Carrying out a symmetry analysis on the remaining 59 structures reveals that in total 238 symmetry inequivalent possible diffusion paths exist. Approximately 53\% of the paths are reflective, making the R-NEB method especially useful to fully characterize more paths while using less computational resources. Even though the R-NEB method improves the speed of relaxing the NEB paths, 238 paths are still extremely computationally expensive. The NEB calculations are on the order of 3-4 magnitudes slower than the unit cell relaxation (Figure \ref{fig:runtime_mgnbo} (a)). Already the relaxation of the vacancy defect structure is 2-3 orders of magnitude slower than the relaxation of the unit cell. Additionally, we do not consider structures with partial occupancies easily leading to more than the maximum number of symmetry inequivalent paths found in the more symmetric materials in this study.\\
Here, we use a subset of the structures which have at most four symmetry inequivalent paths and at least one percolating path consisting of solely reflective paths. In this way, it is possible to characterize more different crystal structures in terms of their ionic diffusivity while obtaining a diverse dataset. We find structures that contain solely non-reflective paths or solely reflective paths, as well as structures that contain both of the aforementioned. An example of a structure containing a non-reflective as well as a reflective percolating path is shown in Figure \ref{fig:runtime_mgnbo} (b).

Due to distortions in the material, we further disregard structures containing Nitrogen as an anion. This process left us with 40 symmetry inequivalent reflective NEB paths obtained from 16 different materials ranging over 10 different spacegroups.
We further classify the different materials into structure groups including: spinel (8 structures), garnet (3), layered (2, Mg$_3$Nb$_6$O$_{11}$ and MgMn$_3$O$_7$), chevrel (1), $P4_2/mbc$ (1, Mg(SbO$_2$)$_2$) and $Pnma$ (1, MgSb$_2$Cl$_2$O$_3$).

\subsection{Structural descriptors}
\begin{figure}[h]
    \centering
    \includegraphics[width=.95\linewidth]{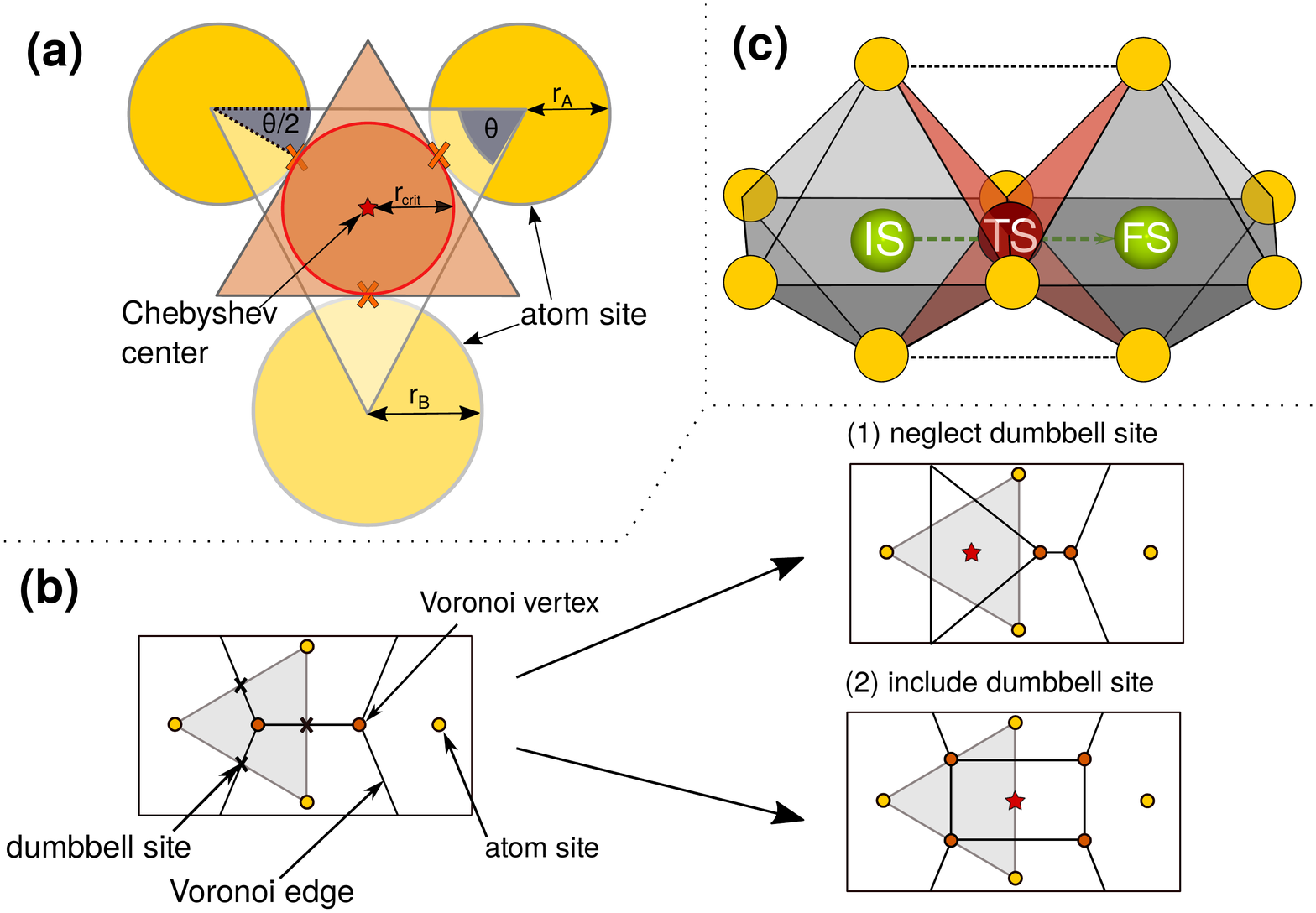}
     \caption{(a) Estimation of the critical radius using the Chebyshev center in a three-fold coordinated face consisting of atoms with the two different radii $r_A$ and $r_B$.  (b) Motivation for including dumbbell sites when finding transition state geometries. Given four atoms arranged in a rhombus like shape, the consideration of only Voronoi vertices leads to identifying three nearest neighbors while including the dumbbell site leads to four nearest neighbors. (c) Dumbbell-shaped transition state (TS) with a possible six-fold coordination indicated by the dark-red planes. Initial (IS) and final state (FS) are in an octahedral coordination.}
     \label{fig:geometrical_explained}
\end{figure}

Structural and chemical descriptors describing the ion mobility in bulk structures have been discussed extensively in literature. \cite{dowty1980crystal, rong2015materials, wang2015design, katcho2019investigation, he2020cavd} Three main contributions to the barrier obtained from DFT calculations have been identified to be (i) electrostatic interactions as well as (ii) electronic and (iii) ionic relaxation effects. \cite{zimmermann2018electrostatic, levi2009review, safran1980electrostatic} Relaxed NEB paths allow to study the effect of electronic and ionic relaxations in the structure. In this work, we investigate simple descriptors that correlate with the NEB-barrier making it possible to group diffusion channels. It is highly desirable to find descriptors that can be directly derived from the relaxed unit cell, since this is the least expensive calculation of the workflow. Rough estimates on whether or not the diffusion path is structurally similar to previously relaxed paths are acceptable in a screening study, as long as the needed descriptors are fast to calculate. In order to describe the transition state automatically before relaxing the NEB path, we introduce a path finder algorithm that predicts the transition state geometry given the stable insertion sites of initial (IS) and final (FS) position of the moving ion in the path as an input (Figure \ref{fig:algorithm} - step (1)). Here, this algorithm is developed for multivalent magnesium cathode materials, but the ideas are general enough to extend to other battery chemistries. The reason for not applying it directly to other cations is that different cations prefer different environment specific coordination numbers. \cite{brown1988factors} From literature it is known that often the critical radius, corresponding to the largest free sphere radius possible along the path is crucial (see also red circle in Figure \ref{fig:geometrical_explained} (a)). \cite{katcho2019investigation, he2020cavd} Following the ideas of connecting Voronoi nodes and edges for diffusion channel analysis \cite{blatov2004voronoi,willems2012algorithms}, the path finder algorithm proceeds as follows (steps visualized exemplary in Figure \ref{fig:algorithm}):

\begin{figure}[h]
    \centering
    \includegraphics[width=.6\linewidth]{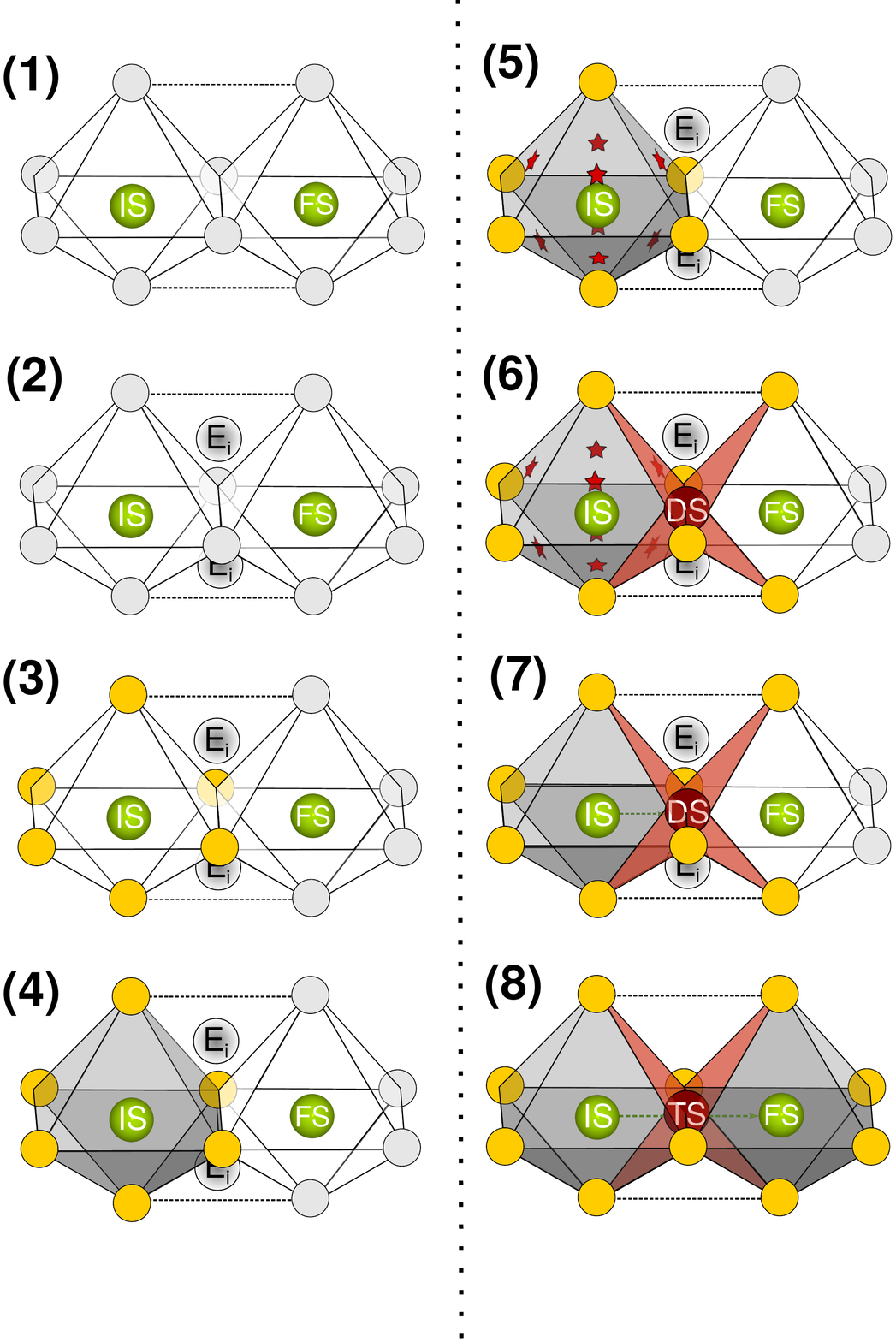}
    \caption{The eight steps of the algorithm visualized on an example structure with two edge-sharing polyhedra. The details of each step are provided in the text.}
    \label{fig:algorithm}
\end{figure}

\begin{enumerate}[label=(\arabic*)]
    \item Start from the user provided initial state (IS) of the path;
    \item Find possible unoccupied ion sites (possible intermediate sites E$_i$) using a Voronoi tessellation of the structure as implemented in pymatgen \cite{ong2013python} including clustering of Voronoi nodes with a tolerance of 1 \si{\angstrom} and removing nodes that are closer than 1 \si{\angstrom} to an atom of the intercalation framework;
    \item Identify all nearest neighbors of the ion at the current position using O'Keeffe's method \cite{o1979proposed} to obtain the Coordination number (CNN) defined as:
    \begin{equation}
        CNN = \sum \frac{\sigma}{\sigma_{max}},
        \label{eq:NN}
    \end{equation}
    where $\sigma$ is the solid angle and $\sigma_{max}$ defines the neighboring atom whose polyhedron site subtends the largest solid angle. Following the recent findings on benchmarking nearest neighbors algorithms by Pan \textit{et. al.}, we consider an atom a neighbor if the solid angle is larger than 50\% of the maximum solid angle, \textit{i.e.} $\sigma > 0.5* \sigma_{max}$. \cite{pan2020benchmarking};
    \item Create faces enclosing the current position;
    \item Investigate each face found in step (4) and find the center. In order to account for possible unequal ionic radii (here we use the Shannon ionic radii \cite{shannon1976revised}) on the edges of the face, we define the center as the Chebyshev center of the face, which is the largest inscribed circle within a polygon (see red star(s) in Figure \ref{fig:geometrical_explained} (a) and Figure \ref{fig:algorithm} - step (5)). For creating the polygon for which the Chebyshev center is calculated, all vertices of the face are considered. From each vertex, a line is drawn from the center of the atom site with an angle $\theta/2$ that interesects with the circle representing the atomic radius (orange crosses in Figure \ref{fig:geometrical_explained} (a)). $\theta$ is the angle spanned by the two edges intersecting at the vertex. The tangent on the ion going through the orange crosses makes up the edges of the inner polygon for which the Chebyshev center is calculated;
    \item Investigate each edge of the face at the center in between the two vertices again considering ionic radii. Find the CNN at this dumbbell site (DS) according to the definition in step (3) (see also image on the left in Figure \ref{fig:geometrical_explained} (b)). If the CNN is larger than the CNN at the Chebyshev center, choose the dumbbell site. In Figure \ref{fig:algorithm} - step (6) this is highlighted by the red planes indicating the six accessible neighbors at this position in contrast to three accessible neighbors at the Chebyshev center;
    \item Consider all valid dumbbell sites and Chebyshev centers and choose the position that minimizes the distance to the final position. In the example in Figure \ref{fig:algorithm} this is a DS.;
    \item Consider the next position, being either an accessible unoccupied site E$_i$ or the final state (FS). If an unoccupied site is accessible, repeat steps (3) - (8) until the final position is reached.
\end{enumerate}

The critical radius along the path is then defined as the minimum radius of a sphere in all possible dumbbell and face positions along the path. Moreover, this position is then denoted as possible transition state configuration (TS in Figure \ref{fig:algorithm} - step (8)). While connecting Voronoi node and edges has been applied successfully in structure and topology characterisation of the void space inside a structure \cite{willems2012algorithms}, here we extend these ideas through considering additional high-symmetry points, namely dumbbell sites. These structural configurations, that can for instance occur for edge-sharing polyhedra, are especially relevant for multivalent ions which favor a larger number of coordinating anions (\textit{e.g.} six-fold for Mg$^{2+}$ \cite{brown1988factors}). Furthermore, the algorithm analyzes only segments of diffusion channels by considering arbitrary initial and final positions provided as an input. This makes it particularly useful for investigating NEB paths for which the initial and final positions are known.\\

Considering the dumbbell site is further motivated through the example shown in Figure \ref{fig:geometrical_explained} (b). The atom sites are arranged in a rhombus. The Voronoi tessellation identifies the Voronoi vertex as a possible transition state of the face under investigation (grey shaded triangle). A secondary Voronoi tessellation around this vertex results in three nearest neighbors (red star in Figure \ref{fig:geometrical_explained} case "(1) neglect dumbbell site"). Applying the Voronoi tessellation around the dumbbell site at the center of the rhombus (red star in Figure \ref{fig:geometrical_explained} (b) case "(2) include dumbbell site") leads to four nearest neighbors being identified. While this case only shows a simplified example in 2D, the approach directly translates into three-dimensional spaces. It allows to identify transition states that can be described by a dumbbell of anions, \textit{i.e.} a dumbbell-hop. \cite{van2001lithium} Such a dumbbell-shaped transition state configuration is shown in Figure \ref{fig:geometrical_explained} (c) for which the algorithm has been described earlier. For instance, we find that the transition state configuration for the Chevrel phase can be best described through a dumbbell site. Additionally, one of the Chevrel paths also has the lowest kinetic barrier observed in this study making the identification of such diffusion characteristics particularly interesting. \\

\subsection{Correlation between geometric descriptor and kinetic barriers}
The path finder algorithm is able to give an estimate of the unrelaxed transition state geometry based on solely the relaxed unit cell. This is important, since it allows to study features of the unrelaxed initial and transition state of the NEB path. For instance, for the case of the spinel structure, the path finder algorithm takes as an input the initial and final positions of the magnesium ion residing in a tetrahedral coordination. As a results, it predicts the path to traverse through a triangular face, into an octahedral unoccupied site. From this intermediate state it subsequently traverses through a second triangular face to the final position of the NEB path. The critical radius is predicted to be found at the triangular faces in agreement with literature. \cite{liu2015spinel, rong2015materials} \\
The considered activation barriers obtained from the CI-NEB calculations for the 40 paths range from 0.23 eV up to 1.42 eV. The lowest barrier is obtained for one of the symmetry inequivalent paths of the Chevrel phase. As mentioned before, the path finder algorithm finds that the transition state is located at a dumbbell site for the Chevrel phase. While the face is three fold coordinated, investigating the dumbbell site leads to four nearest sulfur neighbors being accessible. The fourth nearest neighbor is in fact a coordinating atom of the final site. We note that while this is the only material in the dataset that exhibits this geometrical feature, it coincides with the smallest barrier observed. It is therefore a specific geometrical characteristic that can be exploited automatically through using the presented approach.\\
DFT-NEB calculations allow us to study electronic and ionic relaxation effects on the transition state, which have been found to impact the barrier. \cite{zimmermann2018electrostatic, safran1980electrostatic} As a possible descriptor for studying these effects we use the Root Mean Square Deviation (RMSD) between the first-shell nearest neighbor anionic positions surrounding the migrating ion (anion$_{NN}$ according to equation \ref{eq:NN}) of the initial and relaxed supercell structure upon vacancy defect creation. Soft dynamical modes causing rotation and vibration of nearby anions are expected to directly influence their ability to stabilize the transition state.\cite{kweon2017structural} This feature still circumvents relaxing the full NEB path.

In order to motivate the sub-grouping of diffusion paths, we start by measuring the relationship of the structural distortions of the coordinating anions on the barrier using Pearson correlation coefficients. Separating the spinel structure group (accounting for 24/40 data points) from the remaining five groups leads to a moderate correlation with a Pearson coefficient of +0.37. Magnesium can attract anions more easily in the high vacancy limit leading to a stronger structural distortion. At the same time, the decreased volume upon removing all remaining magnesium ions increases the kinetic barrier for spinel structures explaining the positive correlation. Conversely, the Pearson correlation coefficient for the remaining five groups when plotted against the barrier is -0.24. This trend can be explained by considering for instance the garnet structures, for which the increased ion mobility at the charged states (high vacancy limit) leads to a stabilization of the transition state and therefore lower barriers. In the case of the garnet structures in the charged state, the increased anion mobility can also be directly related to the weak thermodynamic stability, \textit{i.e.} the convex hull energies are above 0.35 eV/atom.\\
When choosing to correlate all NEB barrier values obtained in this work, the Pearson coefficient suggests only a very weak correlation of -0.07 (see Table \ref{tab:pearson} - feature 7). In addition to the dataset being rather small in size, this clearly indicates that it would be beneficial to group the symmetry inequivalent NEB paths. This allows to find descriptor relations within similar path geometries and also paves the way for models which are able to predict the value directly in a screening study.\\

\begingroup
\setlength{\tabcolsep}{10pt} 
\renewcommand{\arraystretch}{1.5} 
\begin{table}[]
\begin{tabular}{lll}
\hline
  & Feature                                                            & \begin{tabular}[c]{@{}l@{}}Pearson correlation\\ coefficient\end{tabular} \\ \hline
1 & Minimum ion-anion$_{NN}$ separation distance initial configuration & -0.09                                                                     \\
2 & Maximum ion-anion$_{NN}$ separation distance initial configuration & +0.02                                                                     \\
3 & Minimum ion-anion$_{NN}$ separation distance transition state      & -0.25                                                                     \\
4 & Maximum ion-anion$_{NN}$ separation distance transition state      & -0.38                                                                     \\
5 & Minimum ion-atom$_{host}$ separation distance transition state     & -0.31                                                                     \\
6 & Change of CNN between initial and transition state                 & +0.31                                                                     \\ \hline
7 & Distortion of anion$_{NN}$ atoms upon vacancy creation          & -0.07     
                                                             
\end{tabular}
\caption{Pearson Correlation coefficients for seven different features measuring the correlation with the kinetic barrier obtained through the NEB-calculations. The anion$_{NN}$ contain all first-shell anionic nearest neighbors obtained using equation \ref{eq:NN}. Features 1 - 6 are considered for the Principal Component Analysis.}
\label{tab:pearson}
\end{table}
\endgroup

\subsection{Sub-grouping diffusion topologies using a principal component analysis}
One approach to visualize possible clusters of similar NEB paths can be obtained through a principal component analysis (PCA). PCA is a technique to find a lower-dimensional representation of a high-dimensional dataset while maximizing the variance captured by the different features (here descriptors) in the dataset. Here, we try to use a minimal set of descriptors given the limited amount of data points. Geometric descriptors have been found to be most relevant for finding fast lithium-ion conductors. \cite{sendek2017holistic} The NEB barrier is calculated as the total energy difference between the initial and transition state configuration. Using the unrelaxed transition state configuration guess (TS) obtained from the path finder algorithm as well as the initial unrelaxed configuration (IS) we extract for both configurations a total of six geometrical descriptors given in Table \ref{tab:pearson} (features 1 - 6). The first four features measure the maximum and minimum distance to all neighboring anions at the TS and IS. While  it is possible to identify all nearest neighbors in the first neighbor shell through employing a Voronoi tessellation, it does not contain information on possible interactions with elements in the second nearest neighbor shell. Therefore, we also include the minimum ion-atom$_{host}$ separation distance where the atom$_{host}$ structure comprises all elements that are not an anion$_{NN}$ or the moving species. Finally, the change of coordination number between the initial and transition configuration is included which has been found to be correlated to the barrier. \cite{liang2020current} \\
In general, we find a moderate correlation between the barrier and the local geometry for the transition state (Features 4+5 with correlations between 0.3 - 0.5) and no correlation between the barrier and the initial state (Features 1+2 with correlations well below 0.3) underlining the importance of describing the transition configuration when trying to estimate the associated kinetic barrier of the path. All six features are readily available and only require the relaxation of the supercell.

\begin{figure}[h]
    \centering
    \includegraphics[width=.75\linewidth]{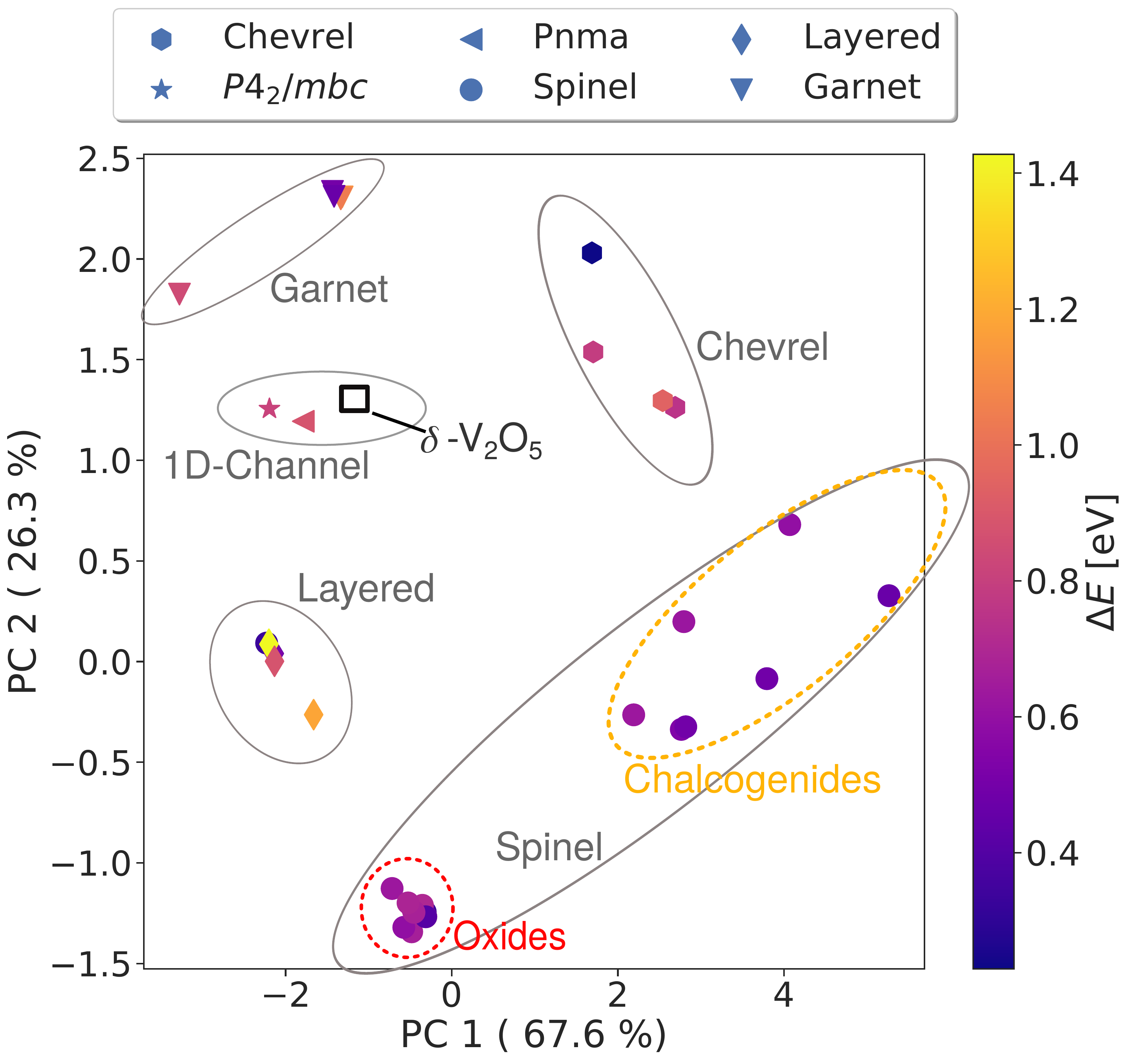}
    \caption{PCA of the six descriptors based on the unrelaxed initial and transition state configuration. $\Delta E$ corresponds to the value of the kinetic barrier obtained from the NEB calculations. The black open square indicates the structure $\delta$-V$_2$O$_5$, which has been discarded during the screening process before estimating kinetic barriers. The spinel compounds are further subdivided into the anion groups oxides (red  dotted circle) and chalcogenides (S, Se, Te, yellow dotted ellipse).}
    \label{fig:pca_analysis}
\end{figure}

Figure \ref{fig:pca_analysis} shows the PCA using the six discussed descriptors for the symmetry inequivalent NEB paths. The two first components of the PCA carry $\sim$ 94\% of the variance in the data meaning it is reasonable to visually inspect the data projected on only two dimensions. The different crystal groups spinel, chevrel, layered and garnet appear well separated. Additionally, the two symmetry inequivalent paths found in the Chevrel structure appear separated. Especially the smallest barrier found in the dataset for the path in the dilute vacancy limit for the Chevrel structure can be distinguished from paths with larger barriers (blue point in group marked as "Chevrel" in Figure \ref{fig:pca_analysis}). The separation of data points for the spinel group can be mainly attributed to the different anions O, S, Se and Te. All spinel compounds containing oxygen are grouped together on the left hand side due to the short Mg-O bond length distances (indicated as red dotted circle in Figure \ref{fig:pca_analysis}). For spinel compounds, larger anions (S, Se and Te - indicated as yellow dotted ellipse in Figure \ref{fig:pca_analysis}) lead to improved ion diffusivity due to an increased critical radius. \cite{canepa2017high} All barriers for the NEB paths in the Spinel group are found to be 0.8 eV at most, while all 24 paths only differ by 0.47 eV. Thus, adding additional data points from other Spinel compounds falling in the same area will only add minor information.\\
An example of how to use this approach to more efficiently find new structures that have not been explored is to look at areas, where data points are sparse or the barrier values vary in between data points close by. The latter indicates that the given descriptors do not capture all information necessary to determine the barrier which will be the topic of a subsequent study. For instance, NEB barriers of the five paths of the group of layered materials differ by 1.0 eV. Nevertheless, we find the material Mg$_3$Nb$_6$O$_{11}$ to have percolating paths (path B-B in Figure \ref{fig:runtime_mgnbo} (b)) with low barriers. In the dilute vacancy limit (discharged) the NEB barrier is 0.42 eV, while at the high vacancy limit (charged) it is 0.53 eV. To fully characterize this structure and its ionic conductivity, all non-reflective percolating paths need to be investigated which is beyond the scope of this study.

In order to test that the approach also works beyond the structures in the dataset, we include the $\delta$-phase of the known cathode material V$_2$O$_5$ \cite{chernova2009layered,amatucci2001investigation} in the PCA. Initially, this structure was removed when choosing the initial pool of materials, since upon linear initialization of the path the moving ion is closer than 1 \si{\angstrom} to an anion along the path. The path finder algorithm finds the transition state configuration at the three-coordinated oxygen face in accordance with literature. \cite{rong2015materials} Additionally, it finds the intermediate square pyramidal site as an accessible unoccupied site. Replacing the linear initialization guess with the found intermediate states from the path finder algorithm can be seen as a viable approach to also find initial non-linear path guesses for the NEB calculations. The added data point of the $\delta$-V$_2$O$_5$ structure in the PCA (open black square in Figure \ref{fig:pca_analysis}) is in close vicinity to Mg(SbO$_2$)$_2$ ($P4_2/mbc$ - 0.81 eV barrier in the discharged state) and  MgSb$_2$Cl$_2$O$_3$ ($Pnma$ - 0.88 eV barrier in the discharged state) that both have 1D-like diffusion channels. Although $\delta$-V$_2$O$_5$ is considered a pseudo-layered material, \cite{rong2015materials} the diffusion channels are 1D-like which matches the diffusion topologies of the data points close by (labeled as 1D-Channel in Figure \ref{fig:pca_analysis}). This underlines the importance of being able to group NEB paths based on unrelaxed structures as it allows to focus on identifying new or specific diffusion topologies (data points in close vicinity to others) or mechanisms enabling fast diffusion in similar topologies (\textit{e.g.} lowered barrier for the garnet structure in the charged state enabled through high anion mobility).

\section{Conclusion}
We have conducted a systematic study of the kinetic properties of magnesium diffusion in bulk cathode materials using the NEB method. 16 different materials were calculated resulting in a dataset containing 40 symmetry inequivalent diffusion paths. The anionic framework of the materials consisted of either O, S, Se, Te or Cl. In order to extract relevant descriptors for the NEB barrier, we introduce a path finder algorithm that takes as an input only the initial and the final position of the moving ion along the path. The algorithm is able to estimate a transition state configuration solely based on the relaxed unit cell which gives a speed-up of 3-4 orders of magnitude compared to calculating the relaxed transition state in screening studies. Investigating the relation between the the distortion of the anionic framework upon vacancy defect creation and the barrier, reveals the importance of  being able to distinguish NEB paths into sub-groups. While the strong distortion of the anionic framework lowers the NEB barrier for garnet structures as a consequence of decreased thermodynamic stability, it is non-beneficial to the diffusivity in the case of the spinel structures. This means, that descriptors of the kinetic barriers found for specific systems, can not always be generalized to  crystal structures outside of the data at hand. This is important when designing filter criteria for screening studies.\\
By extracting six geometrical descriptors from the initial and transition state configuration a PCA was carried out. We find that a PCA can help in identifying sub-groups with similar diffusion topologies automatically. In detail, we find that such an approach can support a more efficient screening study through: (1) Identifying points in the reduced dimensionality space that are in no close vicinity to any other points, indicating a possible new diffusion topology in the dataset; (2) Finding points that are in close vicinity but show significant differences in the observed barriers pointing towards peculiarities in the ion migration. For instance, we find the latter case for the garnet structures. With an increasing amount of data, this also holds the possibility to employ unsupervised learning techniques for sub-grouping diffusion topologies.\\
This work paves the way for more efficient screening studies when searching for fast ionic cathode conductors. This is achieved through employing simple but informative descriptors that can lead to a speed-up on the order of three-magnitudes compared to running DFT calculations in combination with the NEB method. The descriptors are extracted fully automatically using an algorithm that was specifically developed for multivalent cathode materials. The ideas of the algorithm, allowing the identification of dumbbell hops, can in principle directly be transferred to other battery chemistries as well as other applications which require the automatic identification of diffusion channel descriptors.

\section{Computational Methods}
Density Functional Theory calculations were performed within the Perdew-Burke-Ernzerhof revised for solids (PBEsol) functional \cite{csonka2009pbesol, perdew1996generalized}  using the Vienna Ab-initio Simulation Package (VASP). \cite{kresse1996efficient, kresse1999ultrasoft, blochl1994projector} The energy cut-off for the plane wave basis set chosen is 520 eV. To sample the Brillouin zone, we ensure a k-point density $>$ 4.7 /\si{\angstrom}$^{-1}$ and the forces on each atom are converged under 0.03 eV/\si{\angstrom}. In order to reduce the Coulombic self-interaction error we employ a Hubbard $U$-correction \cite{cococcioni2005linear} on the $d$-orbitals of the transition metal in structures containing oxygen (values taken from the materialsproject \cite{ong2013python}):  U$_{V}$ = 3.25 eV, U$_{Co}$ = 3.32 eV, U$_{Cr}$ = 3.7 eV, U$_{Mn}$ = 3.9 eV ,U$_{Mo}$ = 4.38 eV, U$_{Fe}$ = 5.3 eV, U$_{Ni}$ = 6.2 eV.
For estimating the phase stability of the structures, we construct the convex hull from all known stable structures as found in the materialsproject database and calculate the convex hull energy of the phase of interest. \cite{ong2013python} The gas phase reference energy of oxygen at 0K is related to the energy difference between gas phase water and hydrogen molecules including the zero point energy corrections as suggested by Rossmeisl \textit{et al.} \cite{rossmeisl2005electrolysis}.

All NEB-simulations have been carried out using the Atomic Simulation Environment.\cite{larsen2017atomic} The forces acting on the images along the path are converged to less than 0.05 eV/\si{\angstrom}. As discussed by Liu \textit{et al.} \cite{liu2015spinel}, we do not apply a Hubbard correction for the NEB-calculations since the transition state is a less localized state than the initial and final states of the path. Defect-defect interactions are accounted for through assuring a minimum distance of 8 \si{\angstrom} in between repeating supercells.

\section{Acknowledgment}
The authors wish to acknowledge support from the European Magnesium Interactive Battery Community(e-Magic) FET-Proactive project (Contract N. 824066). FTB, TV, IEC acknowledge support from the Department of Energy Conversion and Storage, Technical University of Denmark, through the Special Competence Initiative Autonomous Materials Discovery (AiMade). JMGL acknowledges support from the Villum Foundation’s Young Investigator Programme (4th round, project: In silico design of efficient materials for next generation batteries. Grant number: 10096). IEC acknowledge support from the Independent Research Fund Denmark (Danish ERC Programme, project "Multiscale Design of Electrochemical Metamaterials". Grant number: 0227-00001B).

\section{Keywords}
Electrochemistry, Density functional calculations, Diffusion barriers, Geometrical descriptors, Machine learning

\clearpage 

\section{TOC}

Through applying a path finder algorithm, which automatically extracts geometrical features of the unrelaxed transition state, it is possible to sub-group ionic diffusion channel topologies. This enables to more efficiently explore unknown crystal structures when searching for fast ionic conductors in screening studies.

\begin{figure}[h]
    \centering
    \includegraphics[width=.6\linewidth]{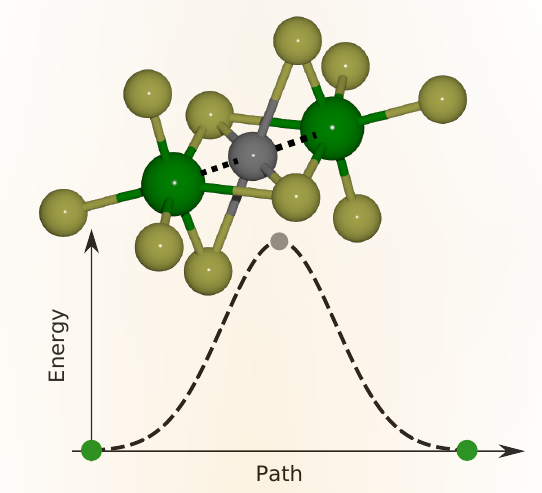}
    \caption{TOC entry graphic.}
    \label{fig:number_neb_paths}
\end{figure}
\clearpage



\end{document}